%% file: fse2021.tex
\documentclass[sigconf,screen]{acmart}
\acmConference[ESEC/FSE 2021]{The 29th ACM Joint European Software Engineering Conference and Symposium on the Foundations of Software Engineering}{23 - 27 August, 2021}{Athens, Greece}

\AtBeginDocument{%
  \providecommand\BibTeX{{%
    \normalfont B\kern-0.5em{\scshape i\kern-0.25em b}\kern-0.8em\TeX}}}

\setcopyright{none}

\newcommand{\tool}{\textsc{PieProf}\xspace}
\newcommand{\ff}{native function call\xspace}
\newcommand{\ffs}{native function calls\xspace}
\newcommand{\yu}[1]{\textcolor{blue}{#1}}
\newcommand{\jtan}[1]{\textcolor{orange}{#1}}

\providecommand{\myparabb}[1]{\smallskip\noindent\textbf{#1} }

\usepackage[normalem]{ulem}
\usepackage{xspace}
\usepackage{listings}
\usepackage{graphicx}
\usepackage{adjustbox}
\usepackage{booktabs} 

\usepackage{algorithm}
\usepackage{amsmath}
\usepackage{algpseudocode}
\usepackage{tabto}
\usepackage{caption}
\usepackage{subfig}
\usepackage{caption}
\usepackage{flushend}
\usepackage{multirow}
\usepackage{multicol}
\usepackage{enumitem}
\usepackage{bm}
\usepackage{ulem}
\usepackage{pifont}
\usepackage{microtype}

\copyrightyear{2021} 
\acmYear{2021} 
\setcopyright{acmlicensed}\acmConference[ESEC/FSE '21]{Proceedings of the 29th ACM Joint European Software Engineering Conference and Symposium on the Foundations of Software Engineering}{August 23--27, 2021}{Athens, Greece}
\acmBooktitle{Proceedings of the 29th ACM Joint European Software Engineering Conference and Symposium on the Foundations of Software Engineering (ESEC/FSE '21), August 23--27, 2021, Athens, Greece}
\acmPrice{15.00}
\acmDOI{10.1145/3468264.3468541}
\acmISBN{978-1-4503-8562-6/21/08}

\begin{document}

\title{Toward Efficient Interactions between Python and Native Libraries}

\author{Jialiang Tan}
\authornote{Both authors contributed equally to this research.}
\authornote{This work is done when Jialiang visits at NCSU.}
\email{jtan02@email.wm.edu}
\author{Yu Chen}
\authornotemark[1]
\email{ychen39@email.wm.edu}
\affiliation{%
  \institution{William \& Mary}
  \city{Williamsburg}
  \state{Virginia}
  \country{USA}
}

\author{Zhenming Liu}
\email{zliu@cs.wm.edu}
\affiliation{%
  \institution{William \& Mary}
  \city{Williamsburg}
  \state{Virginia}
  \country{USA}
}

\author{Bin Ren}
\email{bren@cs.wm.edu}
\affiliation{%
  \institution{William \& Mary}
  \city{Williamsburg}
  \state{Virginia}
  \country{USA}
}

\author{Shuaiwen Leon Song}
\email{shuaiwen.song@sydney.edu.au}
\affiliation{%
  \institution{University of Sydney}
  \city{Sydney}
  \state{}
  \country{Australia}
  \postcode{}}

\author{Xipeng Shen}
\email{xshen5@ncsu.edu}
\affiliation{%
  \institution{North Carolina State University}
  \city{Raleigh}
  \state{North Carolina}
  \country{USA}
  \postcode{}}
  
\author{Xu Liu}
\email{xliu88@ncsu.edu}
\affiliation{%
  \institution{North Carolina State University}
  \city{Raleigh}
  \state{North Carolina}
  \country{USA}
  \postcode{}}

\renewcommand{\shortauthors}{Tan and Chen, et al.}

\begin{abstract}
Python has become a popular programming language because of its excellent programmability. Many modern software packages utilize Python for high-level algorithm design and depend on native libraries written in C/C++/Fortran for efficient computation kernels. Interaction between Python code and native libraries introduces performance losses because of the abstraction lying on the boundary of Python and native libraries. On the one side, Python code, typically run with interpretation, is disjoint from its execution behavior. On the other side, native libraries do not include program semantics to understand algorithm defects.

To understand the interaction inefficiencies, we extensively study a large collection of Python software packages and categorize them according to the root causes of inefficiencies. We extract two inefficiency patterns that are common in interaction inefficiencies. Based on these patterns, we develop \tool{}, a lightweight profiler, to pinpoint interaction inefficiencies in Python applications. The principle of \tool{} is to measure the inefficiencies in the native execution and associate inefficiencies with high-level Python code to provide a holistic view.
Guided by \tool{}, we optimize 17 real-world applications, yielding speedups up to 6.3$\times$ on application level.
\end{abstract}






\maketitle

\input{intro_zliu}

\input{background.tex}

\input{design.tex}

\input{implementation.tex}

\input{evaluation.tex}
\input{conclusion.tex}

\begin{acks}
We thank the anonymous reviewers for their valuable comments. We thank Denys Poshyvanyk for his feedback to the paper. This work is supported in part by NSF grants CNS-2050007, CRII-1755769, OAC-1835821, IIS-2008557, CCF-1703487, CCF-2028850 and CCF-2047516, a Department of Energy (DOE) grant DE-SC0013700. 
\end{acks}

\bibliographystyle{ACM-Reference-Format}
\bibliography{fse2021}

\appendix


\end{document}

%% file: intro_zliu.tex
\section{Introduction}
In recent years, Python has become the most prominent programming language for data modeling and library development, especially in the area of machine learning, thanks to its elegant design that offers high-level abstraction, and its powerful interoperability with native libraries that delivers heavy numeric computations. Decoupling data analysis and modeling logics from operation logics is the singular mechanism guiding the remarkable improvements in developers’ productivity in the past decade. Python enables small teams to build sophisticated model~\cite{meta} that were barely imaginable a few years ago, and enables large teams of modelers and numeric developers to seamlessly collaborate and develop highly influential frameworks such as Tensorflow~\cite{tensorflow2015-whitepaper} and Pytorch~\cite{paszke2017automatic}. 

While high-level languages to articulate business logics and native libraries to deliver efficient computation is not a new paradigm, downstream developers have not always understood the details of native libraries, and have implemented algorithms that interacted poorly with native codes. A well-known example of the \emph{interaction inefficiency} problem occurs when developers, who fail to recognize that certain matrix operations can be vectorized, write significantly slower loop-based solutions. MATLAB and Mathematica can alleviate the problem since these languages usually are locked with a fixed set of native libraries over a long time, and developers can establish simple best practice guidelines to eliminate most interaction inefficiencies (MATLAB contains the command, “try to vectorize whenever possible”).

In the Python ecosystem, native libraries and downstream application codes evolve rapidly so they can interact in numerous and unexpected ways. Therefore, building a list to exhaust all interaction inefficiencies becomes infeasible. We seek a solution that will automatically identify the blocks of Python code that lead to inefficient interactions, through closing the knowledge gap between Python and native code. Existing profiling tools cannot address this issue. Python profiles~\cite{cProfile, guppy3, py-spy, pyflame, pyinstrument, pycallgraph, pprofile, memoryprofiler, austin} cannot step in native code so they do not know  execution details. Native profiling tools~\cite{reinders2005vtune, de2010new, nistor2013toddler, adhianto2010hpctoolkit, chabbi2012deadspy, wen2017redspy, loadspy, wen2018watching} can identify hotspots, which sometimes offer leads to problematic code blocks. But because these tools do not have knowledge about Python code's semantic, they cannot render detailed root cause and thus often make debugging remarkably challenging. 



We propose \tool, the first lightweight, insightful profiler to pinpoint interaction inefficiencies in Python programs. \tool works for production Python software packages running in commodity CPU processors without modifying the software stacks. 
Its backbones algorithmic module is a recently proposed technique based on hardware performance monitoring units (PMUs) and debug registers to efficiently identify redundant memory accesses (hereafter, referred to as CL-algorithm\footnote{Chabbi-Liu Algorithm.}~\cite{wen2018watching, su2019pinpointing}). CL-algorithm intelligently chooses a small collection of memory cells and uses hardware to track accesses to these cells at a fine granularity. For example, when the technique detects two consecutive writes of the same value to the same cell, it determines that the second write is unnecessary, and flags the responsible statement/function for further inspection. The developer can clearly see where a non-opt memory access occurs and why.  The technique already shows its potential for eliminating inefficiencies in monolithic codebases that use one programming language.


\tool leverages the CL-algorithm in a substantially more complex multi-languages environment, in which a dynamic and (predominantly) interpretation-based language Python is used to govern the semantics and native libraries compiled from C, C++, Fortran are used to execute high-performance computation. Doing so requires us to address three major challenges that crosscuts Python and native code. 


At the measurement front, we need to suppress false positives and avoid tracking irrelevant memory operations produced from Python interpreter and Python-native interactions. For example, memory accesses performed by Python interpreters may ``bait'' the CL-algorithm to waste resources (i.e., debug registers) on irrelevant variables such as reference counters. At the infrastructure front, we need to penetrate entire software stacks: it cannot see execution details (i.e, how memory is accessed) with only Python runtime information, or cannot understand program semantics with only native library knowledge. Our main task here is to compactly implement lock-free calling context trees that span both Python code and native libraries, and retain a large amount of information to effectively correlate redundant memory accesses with inefficient interactions. At the memory/safety front, we need to avoid unexpected behaviors and errors caused by Python runtime. For example, Python’s garbage collection (GC) may
reclaim memory that our tool is tracking. So delicate coordination between \tool and Python interpreter is needed to avoid unexpected behaviors and errors.

We note that while most of the downstream applications we examined are machine learning related, \tool is a generic tool that can be used in any codebase that requires Python-native library interactions.

\myparabb{\textbf{Contributions.}}
We make three contributions.
\begin{itemize}
\item We are the first to thoroughly study the interaction inefficiencies between Python codes and native libraries. We categorize the interaction inefficiencies by their root causes.

\item We design and implement \tool, the first profiler to identify interaction inefficiencies and provide intuitive optimization guidance, by carefully stepping through Python runtimes and native binaries. \tool works for production Python software packages in commodity CPU processors without modifying the software stacks.

\item Following the guidance of \tool, we examine a wide range of influential codebases and identify interaction inefficiencies in 17 real-world applications and optimize them for nontrivial speedups.
\end{itemize}


\myparabb{\textbf{Organization.}}
Section~\ref{background} reviews the background and related work. Section~\ref{characterization} characterizes the interaction inefficiencies. Section~\ref{design} describes the design and implementation of \tool. Section~\ref{evaluation} explains the evaluation. Section~\ref{casestudy} presents case studies. Section~\ref{validity} discusses some threats to validity. Section~\ref{conclusions} presents some conclusions.

%% file: background.tex
\section{Background and Related Work}
\label{background}

\subsection{Python Runtime}
\label{sec:prun}

\myparabb{\textbf{Background on Python.}} Python is an interpreted language with dynamic features. When running a Python application, the interpreter translates Python source code into stack-based bytecode and executes it on the Python virtual machine (PVM), which varies implementations such as CPython~\cite{cpython}, Jython~\cite{jpython}, Intel Python~\cite{intelpython} and PyPy~\cite{pypy}. This work focuses on CPython because it is \emph{the  reference implementation}~\cite{pyimplementation}, while
the proposed techniques are generally applicable to other Python implementations as well. 
The CPython PVM maintains the execution call stack that consists of a chain of {\tt PyFrame} objects known as function frames.  Each {\tt PyFrame} object includes the executing context of corresponding function call, such as local variables, last call instruction, source code file, and current executing code line, which can be leveraged by performance or debugging tools.


\sloppy
Python supports multi-threaded programming, where each Python thread has an individual call stack. Because of the global interpreter lock (GIL)~\cite{gil}, the concurrent execution of Python threads is emulated as regular switching threads by the interpreter, i.e., for one interpreter instance, only one Python thread is allowed to execute at a time.

\begin{figure}[t]
	\centering
	\includegraphics[width=0.85\linewidth]{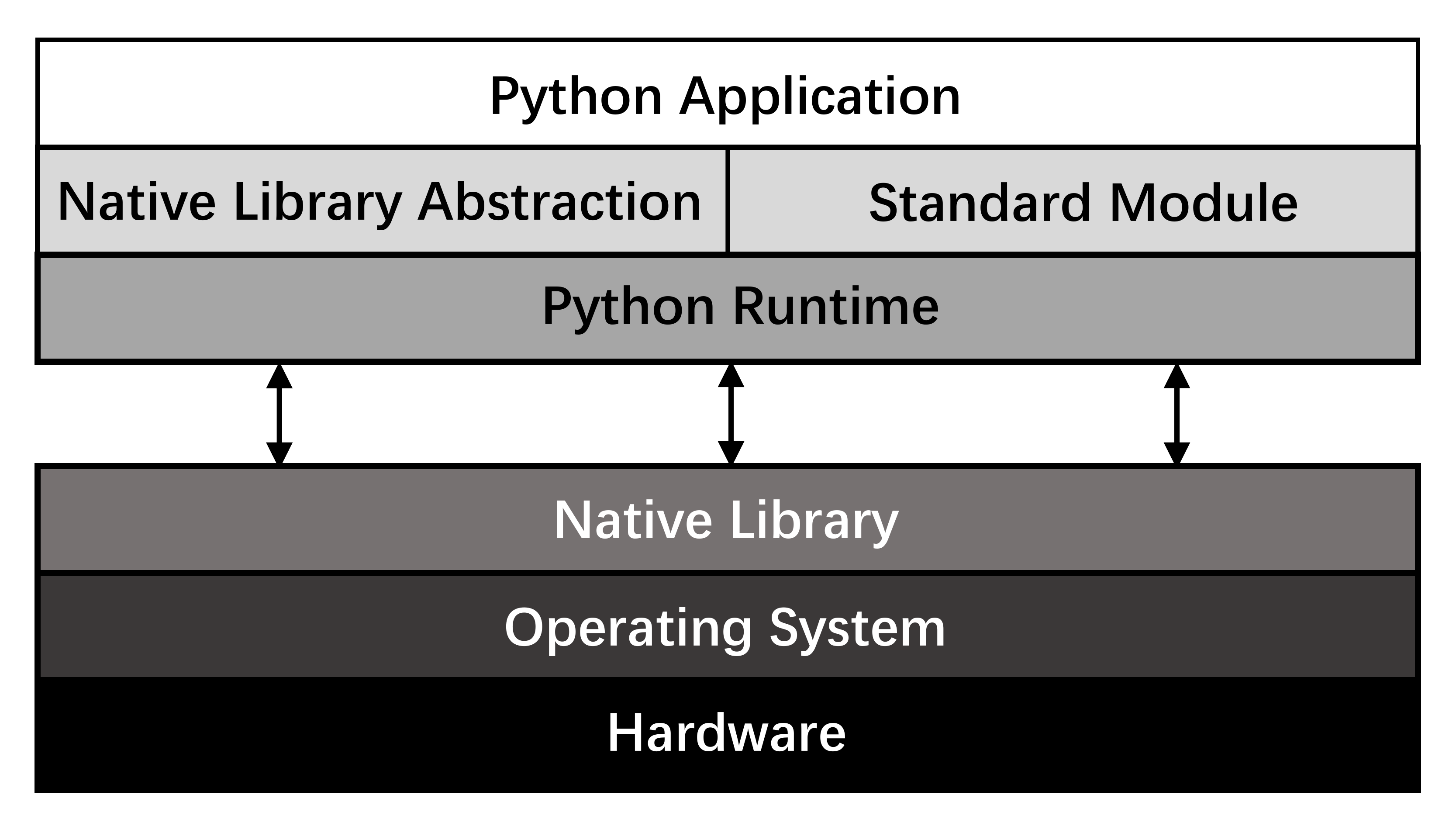}
	\caption{The typical stack of production Python software packages. Python applications usually rely on native libraries for high performance but introduce an abstraction across the boundary of Python runtime and native libraries.}
	\label{fig:hybridmode}
\end{figure}

\myparabb{\textbf{Interaction with native libraries.}}
When heavy-lifting computation is needed, Python applications usually integrate native libraries written in C/C++/Fortran for computation kernels, as shown in Figure~\ref{fig:hybridmode}. Such libraries include Numpy~\cite{van2011numpy, harris2020array}, Scikit-learn~\cite{scikit-learn}, Tensorflow~\cite{tensorflow2015-whitepaper}, and PyTorch~\cite{paszke2017automatic}. Therefore, modern software packages enjoy the benefit from the simplicity and flexibility of Python and native library performance. When the Python runtime calls a native function, it passes the {\tt PyObject}\footnote{{\tt PyObject} is the super class of all objects in Python.} or its subclass objects to the native function. The Python runtime treats the native functions as blackboxes --- the Python code is blocked from execution until the native function returns.

Figure~\ref{fig:hybridmode} shows an abstraction across the boundary of Python runtime and native library, which logically splits the entire software stack. On the upper level, Python applications are disjoint from their execution behaviors because Python runtime (e.g., interpreter and GC) hides most of the execution details. On the lower level, the native libraries lose most program semantic information. This knowledge gap leads to interaction inefficiencies.

\subsection{Existing Tools vs. \tool}\label{sec:related}


This section compares existing tools that analyze inefficiencies in Python and native codes to distinguish \tool.

\myparabb{\textbf{Python performance analysis tools.}}
PyExZ3~\cite{irlbeck2015deconstructing}, PySym~\cite{pysym}, flake8~\cite{flake8}, and Frosted~\cite{frosted} analyze Python source code and employ multiple heuristics to identify code issues statically~\cite{gulabovska2019survey}. XLA~\cite{xla2017xla} and TVM~\cite{chen2018tvm} apply compiler techniques to optimize deep learning applications. Harp~\cite{zhou2020harp} detects inefficiencies in Tensorflow and PyTorch applications based on computation graphs. All of these approaches, however, ignore Python dynamic behavior, omitting optimization opportunities.


Dynamic profilers are a complementary approach. cProfile~\cite{cProfile} measures Python code execution, which provides the frequency/time executions of specific code regions. Guppy~\cite{guppy3} employs object-centric profiling, which associates metrics such as allocation frequency, allocation size, and cumulative memory consumption with each Python object. PyInstrument~\cite{pyinstrument} and Austin~\cite{austin} capture Python call stack frames periodically to identify executing/memory hotspots in Python code. PySpy~\cite{py-spy} is able to attach to a Python process and pinpoint function hotspots in real time. Unlike \tool, these profilers mainly focus on Python codes, with no insights into the native libraries.

Closely related to \tool{}, Scalene~\cite{berger2020scalene} separately attributes Python/native executing time and memory consumption. However, it does not distinguish useful/wasteful resources usage as \tool does. 




\myparabb{Native performance analysis tools.} 
While there are many native profiling tools~\cite{reinders2005vtune, de2010new, adhianto2010hpctoolkit}, from which the most related to Python that can identify performance inefficiencies are Toddler~\cite{nistor2013toddler} that identifies redundant memory loads across loop iterations, and LDoctor~\cite{song2017performance} that reduces Toddler's overhead by applying dynamic sampling and static analysis. DeadSpy~\cite{chabbi2012deadspy}, RedSpy~\cite{wen2017redspy}, and LoadSpy~\cite{loadspy} analyze dynamic instructions in the entire program execution to detect useless computations or data movements. Unfortunately, all of them use heavyweight binary instrumentation, which results in high measurement overhead, and they do not work directly on Python programs.

\subsection{Performance Monitoring Units and Hardware Debug Registers}
Hardware performance monitoring units (PMUs) are widely equipped on the modern x86 CPU architectures. Software can use PMUs to count various hardware events like CPU cycles, cache misses, et cetera. Beside the counting mode that counts the total number of events, PMUs can be configured in sampling, which periodically sample a hardware event and record event's detailed information. PMUs trigger an overflow interrupt when the sample number reaches a threshold. The profiler runtime captures interrupts as signals and collects samples with their executing contexts.

For memory-related hardware events such as memory load and store, Precise Event-Based Sampling (PEBS)~\cite{pebs} in Intel processors provides the effective address and the precise instruction pointer for each sample. Instruction-Based Sampling (IBS)~\cite{ibs} in the AMD processors and Marked Events (MRK)~\cite{srinivas2011ibm} in PowerPC support similar functionalities.

Hardware debug registers~\cite{johnson1982some, mclear1982guidelines} trap the CPU execution when the program counter (PC) reaches an address (breakpoint) or an instruction accesses a designated address (watchpoint). One can configure the trap conditions with different accessing addresses, widths and types. The number of hardware debug registers is limited (e.g., the modern x86 processor has four debug registers).

%% file: design.tex
\section{Interaction Inefficiency Characterization}
\label{characterization}

This section provides a high-level preview of the key findings from applying \tool to an extensive collection of high-profile Python libraries at Github. We specifically categorize the interaction inefficiencies according to the root causes and summarize the common patterns, which serve three purposes: \emph{(i)} this is the first characterization of interaction inefficiencies based on large scale studies, thus rendering a more complete landscape of potential code quality issues that exist in Python codebase for ML and beyond, \emph{(ii)} we see a diverse set of inefficiencies hiding deep in Python-native library interaction, which justifies using heavy machineries/profiling tools to automatically identify them, and \emph{(iii)} these concrete examples explain the common patterns we use to drive the \tool’s design.

\subsection{Interaction Inefficiency Categorization}
\label{classification}

We categorize interaction inefficiencies into five groups. For each category, we give a real example, analyze the root causes, and provide a fix. 



\myparabb{\textbf{Slice underutilization.}}
Listing~\ref{lst:iris1} is an example code from IrisData~\cite{irisdata}, a back-propagation algorithm implementation on Iris Dataset~\cite{fisher1936use}. A loop iterates two multidimensional arrays {\tt ihGrads} and {\tt ihWeights} with indices {\tt i} and {\tt j}  for computation. 
Because Python arrays are supported by native libraries such as Numpy and PyTorch/TensorFlow, indexing operations (i.e., {\tt []}) in a loop trigger native function calls that repeat boundary and type checks~\cite{generalfunction}. 


The so-called vectorization/slicing eliminates repeated ``housework'' and (usually) enables the underlying BLAS~\cite{blackford2002updated} library to perform multi-core computation. Listing~\ref{lst:iris2} shows a simple fix in a 2$\times$ speedup for the entire program execution.

\begin{figure}[t]
\begin{lstlisting}[caption={Interaction inefficiencies in IrisData due to the iteration on Numpy arrays within a loop.},label=lst:iris1]
def train(self, trainData, maxEpochs, learnRate):
    ...
    for j in range(self.nh):
        delta = -1.0 * learnRate * ihGrads[i,j]
        self.ihWeights[i, j] += delta
    ...
\end{lstlisting}
\end{figure}

\begin{figure}[t]
\begin{lstlisting}[caption={Optimized IrisData code with slice notation.},label=lst:iris2]
def train(self, trainData, maxEpochs, learnRate):
    ...
    self.ihWeights[i, 0:self.nh] += -1.0 * learnRate * ihGrads[i, 0:self.nh]
    ...
\end{lstlisting}
\end{figure}

\begin{figure}[t]
\begin{lstlisting}[caption={Interaction inefficiencies in Matplotlib due to the same input {\tt theta}.},label=lst:matplot1]
def rotate(self, theta):
    a = np.cos(theta)
    b = np.sin(theta)
    rotate_mtx = np.array([[a, -b, 0.0], [b, a, 0.0], [0.0, 0.0, 1.0]], float)
    self._mtx = np.dot(rotate_mtx, self._mtx)
    ...
\end{lstlisting}
\end{figure}

\myparabb{\textbf{Repeated native function calls with the same arguments.}}
Functions from native libraries typically have no side effects, so applying the same arguments to a native function results in the same return value, which introduces redundant computations. 
Listing~\ref{lst:matplot1} shows a code from Matplotlib~\cite{Hunter:2007}, a comprehensive library for visualization and image manipulation. This code rotates an image and is often invoked in training neural nets for images. 

The argument {\tt theta} for the {\tt rotate} function (rotate angle) is usually the same across consecutive invocations from deep learning training algorithms because they rotate images in the same batch in the same way. Here, {\tt Pyobject}s returned from native functions {\tt np.cos()}, {\tt np.sin()} and {\tt np.array()} in lines 2-4 have the same values across images that share the same input {\tt theta}. 

This can be fixed by either a simple caching trick~\cite{della2015performance, nguyen2013cachetor}, or refactoring the {\tt rotate} funcion so that it can take a batch of images. We gain a 2.8$\times$ speedup after the fix.

\myparabb{\textbf{Inefficient algorithms.}}
Listing~\ref{lst:scikit1} is an example of algorithmic inefficiencies from Scikit-learn, a widely used machine learning package. 
The code works on {\tt X}, a two-dimensional Numpy array. It calls the native function {\tt swap} from the BLAS library to exchange two adjacent vectors. In each iteration, {\tt swap} returns two {\tt PyObject}s and Python runtime assigns these two {\tt PyObject}s to {\tt X.T[i]} and {\tt X.T[i+1]}, respectively. The loop uses {\tt swap} to move the first element in the range to the end position. Inefficiencies occur because it requires multiple iterations to move {\tt X.T[i]} to the final location.


Instead of using {\tt swap}, we directly move each element to the target location. We apply a similar optimization to the {\tt indices} array as well. Our improvement yields a  6.1$\times$ speedup to the {\tt lars\_path} function.




\begin{figure}[t]
\begin{lstlisting}[caption={Interaction inefficiencies in Scikit-learn due to the inefficient algorithm.},label=lst:scikit1]
def lars_path(X, y, Xy=None, ...):
    ...
    for i in range(ii, n_active):
        X.T[i], X.T[i + 1] = swap(X.T[i], X.T[i + 1])
        indices[i], indices[i + 1] = indices[i + 1], indices[i]
    ...
\end{lstlisting}
\end{figure}

\begin{figure}[t]
\begin{lstlisting}[caption={
Interaction inefficiencies in Metaheuristic~\cite{nguyen2019building, nguyen2018resource} due to the API misuse in native Libraries.},label=lst:motivated1]
def CEC_4(solution=None, problem_size=None, shift=0):
    ...
    for i in range(dim - 1):
        res += 100 * np.square(x[i]**2-x[i+1]) + np.square(x[i]-1)
    ...
\end{lstlisting}
\end{figure}

\myparabb{\textbf{API misuse in native libraries.}}
Listing~\ref{lst:motivated1} is an example of API misuse from Metaheuristic~\cite{nguyen2019building, nguyen2018resource}, which implements the-state-of-the-art meta-heuristic algorithms. The code accumulates the computation results to {\tt res}. Since the computation is based on Numpy arrays, the accumulation operation triggers one native function call in each iteration, resulting in many inefficiencies.

In Listing~\ref{lst:motivated2} shows our fix (i.e., use the efficient {\tt sum} API from Numpy) which avoids most of the native function invocations by directly operating on the Numpy arrays. This optimization removes most of interaction inefficiencies, and yields a 1.9$\times$ speedup to the entire program.

\myparabb{\textbf{Loop-invariant computation.}}
Listing~\ref{lst:loop2} is a code snippet from Deep Dictionary Learning~\cite{mahdizadehaghdam2019deep}, which seeks multiple dictionaries at different image scales to capture complementary coherent characteristics implemented with TensorFlow. Lines 1-3 indicate the computation inputs {\tt A}, {\tt D,} and {\tt X}. Lines 4-5 define the main computation. Lines 6-7 execute the computation with the actual parameters {\tt D\_} and {\tt X\_}. The following pseudo-code shows the implementation:

\qquad \qquad \qquad {\bf for} \ {\it i $\leftarrow$ 1}  to  {\it Iter} \ {\bf do}\\
\centerline{$A = D(X - D^T A)$ }
where {\tt D} and {\tt X} are loop invariants. If we expand the computation, $DX$ and $DD^T$ can be computed outside the loop and reused among iterations, shown as pseudo-code:

\qquad \qquad \qquad $t_1 = DX$

\qquad \qquad \qquad $t_2 = DD^T$

\qquad \qquad \qquad {\bf for} \ {\it i $\leftarrow$ 1}  to  {\it Iter} \ {\bf do}

\centerline{$A = t_1 - t_2 A$ }

This optimization yields a 3$\times$ speedup to the entire program~\cite{zhou2020harp}.

\begin{table}[b]
    \centering
    \scriptsize
    \resizebox{\linewidth}{!}{
    \begin{tabular}{|c|c|}
    \hline
        Inefficiency Pattern & Inefficiency Category  \\
    \hline
    \hline
        \multirow{3}{4em}{Redundant Loads}  & Slice underutilization  \\ 
        \cline{2-2}
        & Inefficient algorithms \\ 
        \cline{2-2}
        & API misuse in native libraries \\ 
    \hline
        \multirow{4}{4em}{Redundant Stores}  & Loop-invariant computation  \\ 
        \cline{2-2}
        & Repeated native function calls with same arguments \\
        \cline{2-2}
        & Inefficient algorithms \\ 
        \cline{2-2}
        & API misuse in native libraries \\
    \hline 
    \end{tabular}
    }
    \caption{Redundant loads and stores detect different categories of interaction inefficiencies.}
    \label{tab:patterns}
\end{table}

\subsection{Common Patterns in Interaction Inefficiencies}
\label{commonpatterns}

We are now ready to explain the common patterns in code that exhibits interaction efficiencies, which we use to drive the design of \tool. Specifically,  we find that almost all interaction inefficiencies involve \emph{(i)} repeatedly reading the same {\tt PyObject}s of the same values, and \emph{(ii)} repeatedly returning {\tt PyObject}s of the same values.

Both observations require developing a tool to identify redundant {\tt PyObject}s, which is difficult and costly because it requires heavyweight Python instrumentation and modification to Python runtime. Further analysis, however, finds that {\tt PyObject} redundancies reveal the following two low-level patterns during the execution from the hardware perspective.
\begin{itemize}[leftmargin=*]
    \item {\em Redundant loads:} If two adjacent native function calls read the same value from the same memory location, the second native function call triggers a redundant (memory) load. Repeatedly reading {\tt PyObject} of the same value result in redundant loads.
    \item {\em Redundant stores:} If two adjacent native function calls write the same value to the same memory location, the second native function call triggers a redundant (memory) store. Repeatedly returning {\tt PyObject} of the same value result in redundant stores.
\end{itemize}
We use the redundant loads and stores to serve as indicators of interaction inefficiencies. Table~\ref{tab:patterns} shows different categories of interaction inefficiencies, which show up as redundant loads or stores. Section 4 describes how we use the indicators. 

\begin{figure}[t]
\begin{lstlisting}[caption={Optimized  Metaheuritics code for Listing~\ref{lst:motivated1}, with appropriate native library API.},label=lst:motivated2]
def CEC_4(solution=None, problem_size=None, shift=0):
    ...
    res += np.sum(100 * np.square(x[0:dim-1]**2 -  x[1:dim]) + np.square(x[0:dim-1] - 1))
    ...
\end{lstlisting}
\end{figure}

\begin{figure}[t]
\begin{lstlisting}[caption={Interaction inefficiencies in Deep Dictionary Learning~\cite{mahdizadehaghdam2019deep} due to loop-invariant computation.},label=lst:loop2]
A = tf.Variable(tf.zeros(shape=[N, N]), dtype=tf.float32)
D = tf.placeholder(shape=[N, N], dtype=tf.float32)
X = tf.placeholder(shape=[N, N], dtype=tf.float32)
R = tf.matmul(D, tf.subtract(X, tf.matmul(tf.transpose(D), A)))
L = tf.assign (A, R)
for i in range(Iter):
    result = sess.run(L, feed_dict={D: D_, X: X_})
\end{lstlisting}
\end{figure}

%% file: implementation.tex
\section{Design and Implementation}
\label{design}

\subsection{Overview}
\begin{figure}[t]
	\centering
	\includegraphics[width=\linewidth]{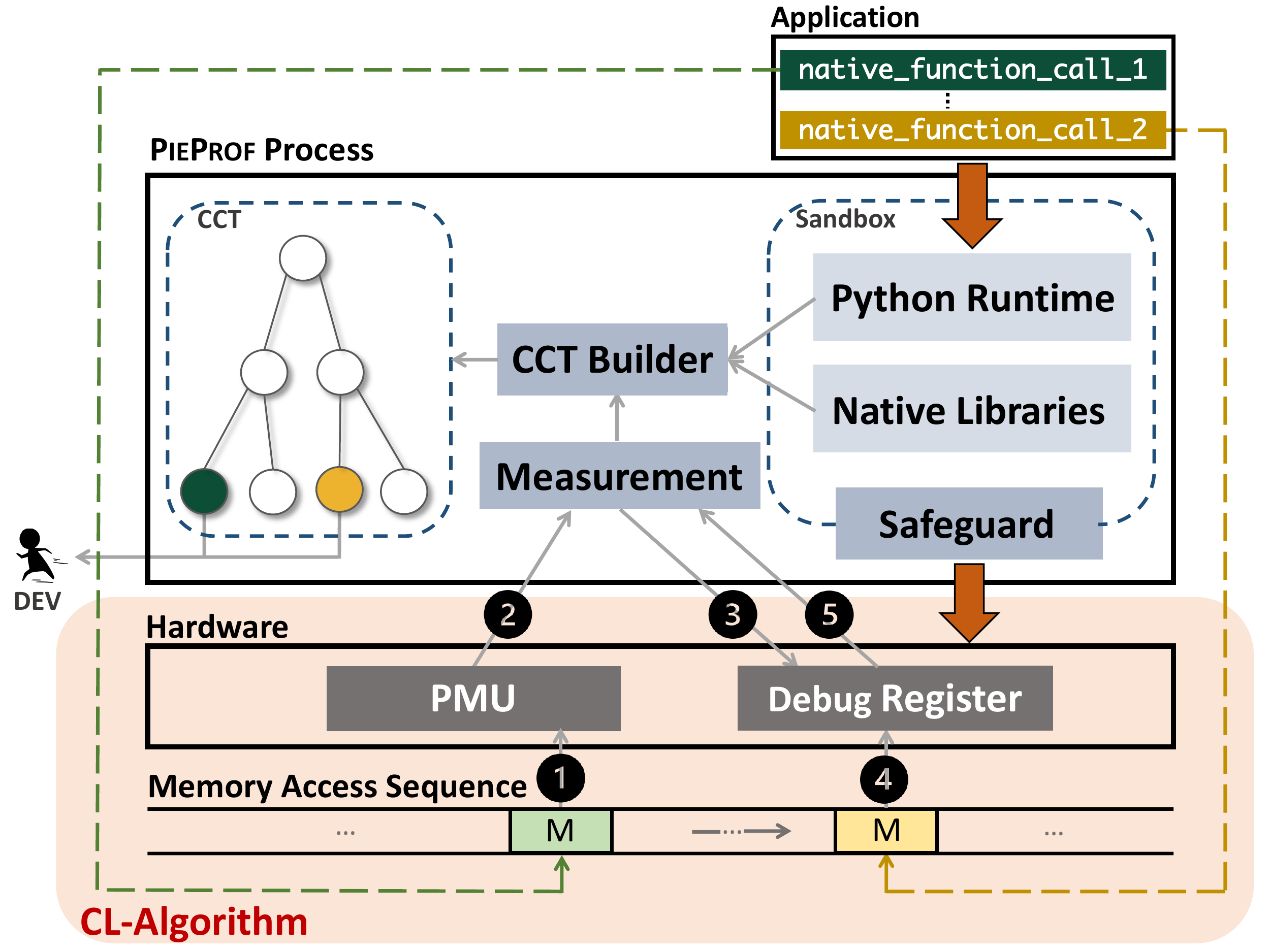}
	\caption{Overview of \tool's workflow}
	\label{fig:new1}
\end{figure}

See Figure~\ref{fig:new1}. Recall that the CL-algorithm controls PMUs and debug registers to report redundant member accesses of a process. \tool interact with Python runtime, native libraries, and the CL-algorithm through three major components: \emph{(i) Safeguard and sandbox.} A thin sandbox is built around Python interpreter and native libraries, and a safeguard is implemented inside the sandbox to moderate communication between Python runtime and the CL-algorithm. \emph{(ii) Measurement.} Upon receiving an event from the CL-algorithm, the measurement component determines whether to notify CCT (calling context tree) builder to update the CCT, and \emph{(iii) CCT Builder.} Upon receiving an update from the measurement component, CCT builder examines Python runtime and native call stacks to update CCT. 

When an interaction inefficiency is detected, it will report to the end user (developer). 

The measurement component helps to suppress false positive and avoid tracking irrelevant variables (e.g., reference counters), the CCT builder continuously update the lock-free CCT, and Safeguard/sandbox ensures that the Python application can be executed without unexpected errors. 

We next discuss each component in details. 

\begin{figure*}[th]
	\centering
	\includegraphics[width=0.725\linewidth]{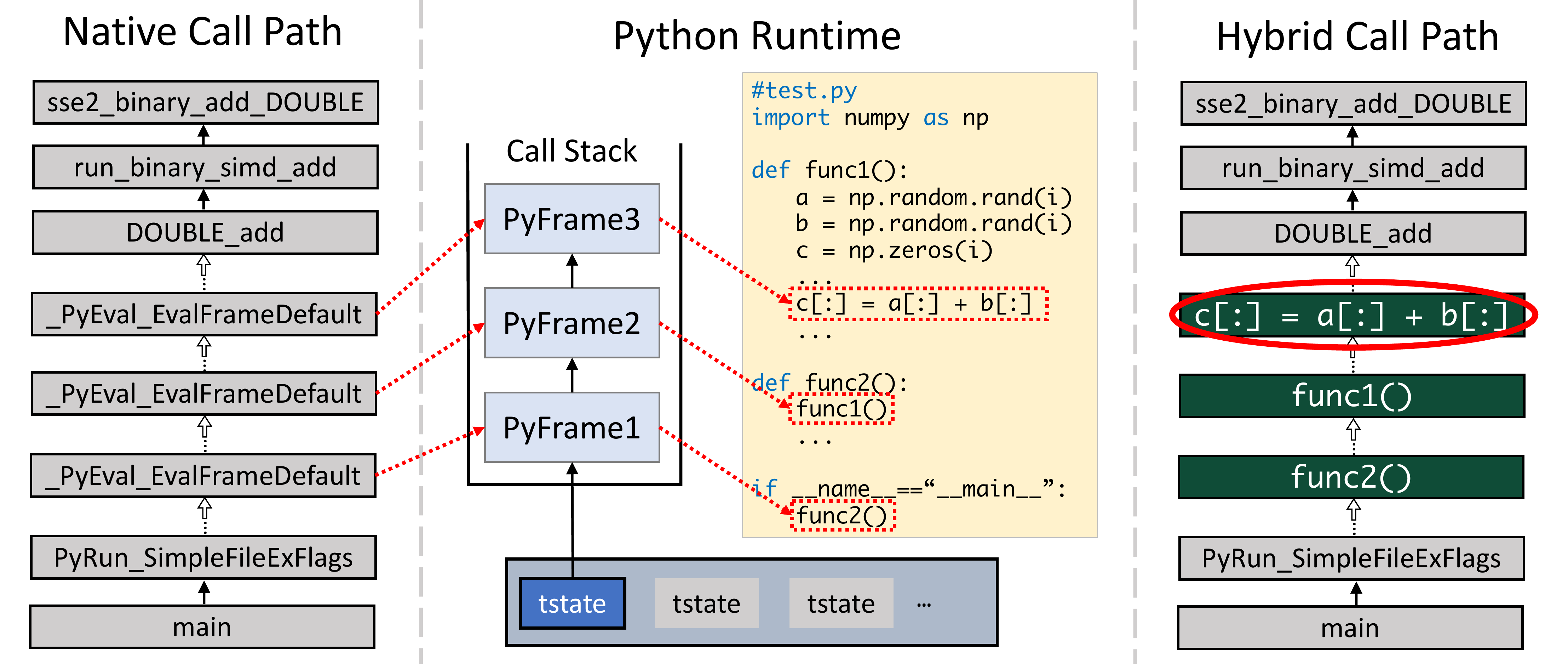}
	\caption{Constructing a hybrid call path across Python runtime and native libraries. White arrows in call paths denote a series of elided call frames in PVM. The red circle in the hybrid call path shows the boundary of Python and native frames, where interaction inefficiencies occur.}
	\label{fig:callpath}
\end{figure*}

\subsection{Measurement}
\label{measurement}
\myparabb{\textbf{CL-algorithm.}}
CL-algorithm uses PMUs and debug registers to identify redundant loads and stores in an instruction stream. It implements a conceptually simple and elegant process: a sequence $a_1, a_2, \dots , a_m$ memory access instructions arrive at the CL-algorithm in a streaming fashion. Here, $a_i$ refers to the address of the memory access for the $i$-th instruction. Upon seeing a new memory access instruction $a_i$ (step 1, i.e \ding{182} in Figure~\ref{fig:new1}), the CL-algorithm uses PMUs to probabilistically determine whether it needs to be tracked (step 2), and if so, store the address in a debug register (step 3). If the debug registers are all used, a random one will be freed up. When a subsequent access to $a_i$ (or any addresses tracked by debug registers) occurs (step 4), the debug register will trigger an interrupt so that the CL-algorithm can determine whether the access is redundant (step 5), by using the rules outlined in 
Section~\ref{commonpatterns}. Since the number of debug registers is usually limited, the CL-algorithm uses a reservoir sampling~\cite{vitter1985random} technique to ensure that each instruction (and its associated memory accesses) has a uniform probability of being sampled.


\myparabb{\textbf{Improving measurement efficiencies.}}
First, PMUs sample instructions at the hardware level so it cannot distinguish memory accesses from the Python interpreter from those from the Python applications. In practice, a large fraction of memory access sequences are related to updating reference counters for Python objects. Therefore, most debug registers will be used to track reference counters if we bluntly use the CL-algorithm, and substantially reduces the chances of identifying memory access redundancies. Second, it needs to ignore redundant memory accesses occurring within the same native function call, or within a code region of \tool because they are not related to interaction inefficiencies. Note that tracking redundant memory accesses within the same native function call is worse than merely producing false positives because it can bury true instances. For example, two write instructions $w_1$ and $w_2$ of the same value are performed on the same memory from function $F_a$, and later function $F_b$ performs a third write instruction $w_3$ of the same value on the same location. If we track redundant accesses within the same function, the CL-algorithm says it has found a redundant pair $\langle w_1, w_2 \rangle$, evicts $w_1$ from the debug register. and never detects the redundant pair $\langle w_1, w_3 \rangle$ caused by the real interaction inefficiencies.


\tool performs instruction-based filter to drop a sample if \emph{(i)} its instruction pointer falls in the code region unrelated to native function calls (e.g., that of \tool), \emph{(ii)} its memory access address belongs to ``junky'' range, such as the head of {\tt PyObject} that contains the reference number. In addition, when the CL-algorithm delivers a redundant memory access pair to \tool, it checks the Python runtime states and drops the sample when these two memory accesses occur inside a same state (corresponding to within the same native function call).

\subsection{Calling Context Trees Builder}
This section first explains the construction of call paths, and then explains how they can be used to construct signal-free calling context trees (CCTs). 


\myparabb{\textbf{Hybrid call path.}}
\tool{} uses libunwind~\cite{libunwind} to unwind the native call path of a Python process to obtain a chain of procedure frames on the call stack. See the chain of ``Native Call Path'' on the left in Figure~\ref{fig:callpath}.
Here, call stack unwinding is not directly applicable to Python code because of the abstraction introduced by PVM. 
The frames on the stack are from PVM, not Python codes. 
For example, the bottom {\tt \_PyEval\_EvalFrameDefault}\footnote{{\tt \_PyEval\_EvalFrameDefault} is a frame (i.e., a function pointer) in the native call stack in runtime that corresponds to invocation of a function or a line of code in Python.} shows up in ``Native Call Path'', but we need the call to correspond to {\tt func2()} in Python code (connected through {\tt PyFrame1}). Thus, \tool{} needs to inspect the dynamic runtime to map native calls with Python calls on the fly. 


\noindent{\emph{1. Mapping {\tt PyFrame} to Python calls.}} First, we observe that each Python thread maintains its call stacks in a thread local object {\tt PyThreadState} (i.e., {\tt tstates} in Figure~\ref{fig:callpath}).  
To obtain Python's calling context, \tool{} first calls {\tt GetThisThreadState()}\footnote{{\tt GetThisThreadState()} is a PVM API to retrieve an object that contains the state of current thread.} to get the {\tt PyThreadState} object of the current thread. Second \tool{} obtains the bottom {\tt PyFrame} object (corresponding to the most recently called function) in the PVM call stack from the {\tt PyThreadState} object. All {\tt PyFrame} objects in the PVM call stack are organized as a singly linked list so we may obtain the entire call stack by traversing from the bottom {\tt PyFrame}.  
Each {\tt PyFrame} object contains rich information about the current Python frame, such as source code files and line numbers that \tool{} can use to correlate a {\tt PyFrame} to a Python method. In Figure~\ref{fig:callpath}, {\tt PyFrame1}, {\tt PyFrame2}, and {\tt PyFrame3} are for Python methods {\tt main}, {\tt func2}, and {\tt func1}, respectively.

 
 \noindent{\emph{2. Extracting {\tt PyFrame}'s from Native Call Path.}} Each Python function call leaves a footprint of {\tt \_PyEval\_EvalFrameDefault} in the native call stack so we need only examine 
 {\tt \_PyEval\_EvalFrameDefault}. Each {\tt \_PyEval\_EvalFrameDefault} maps to a unique {\tt PyFrame} in the call stack of the active thread in Python Runtime. In addition, the ordering preserves, e.g., the third {\tt \_PyEval\_EvalFrameDefault} in ``Native Call Path'' corresponds to the third {\tt PyFrame} in Python's call stack. Therefor use standard Python interpreter APIs to obtain the {\tt PyFrame}'s and map them back to nodes in the native call path.

\begin{figure}[t]
	\centering
	\includegraphics[width=\linewidth]{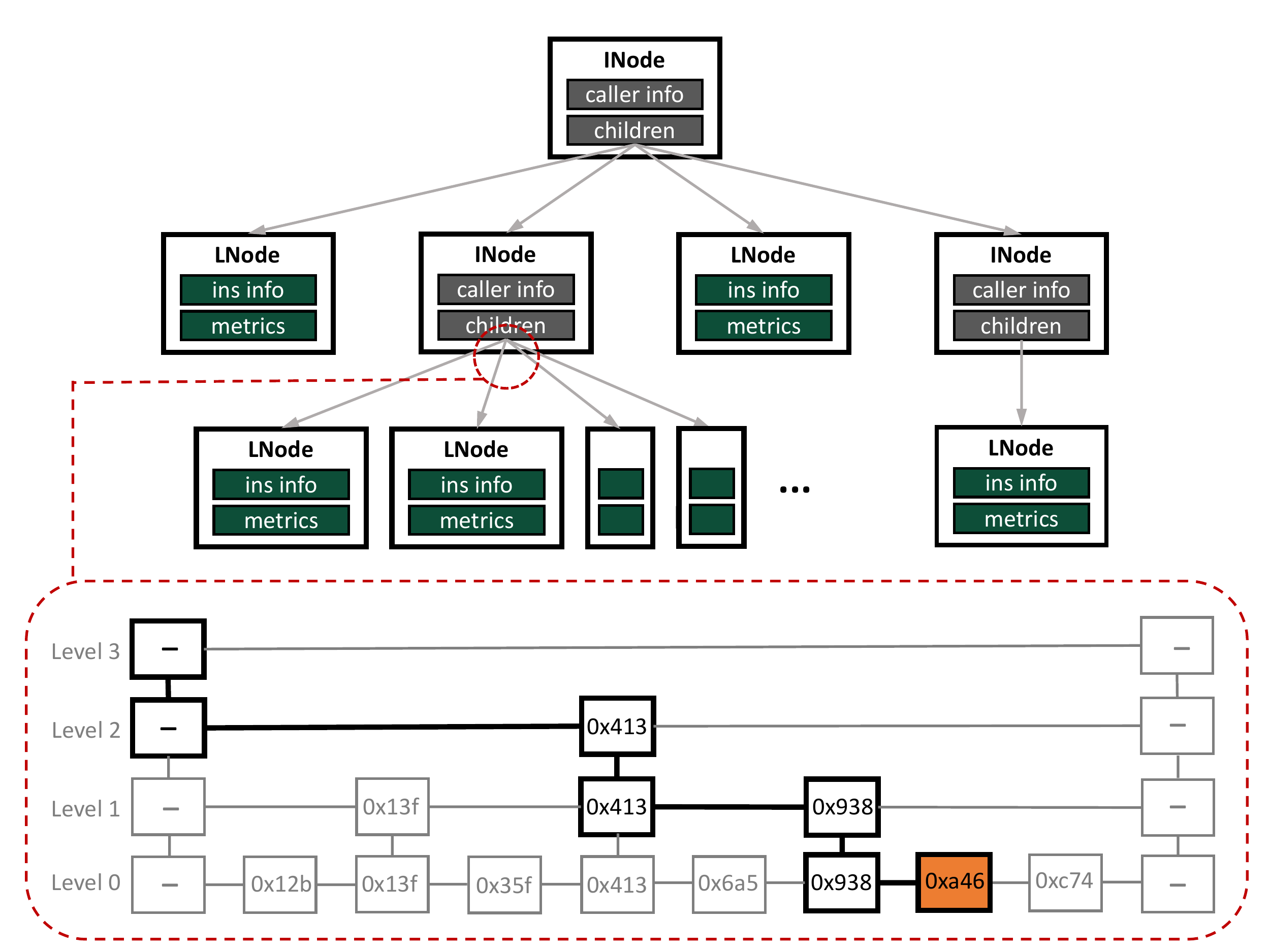}
	\caption{A calling context tree constructed by \tool. Each parent node applies skip-list to organize children. {\tt INode} denotes an internal node and {\tt LNode} denotes a leaf node. Red box shows searching {\tt 0xa46} in the example skip-list.}
	\label{fig:cct}
\end{figure}

\myparabb{\textbf{CCT from call paths.}}
\tool applies a compact CCT~\cite{arnold2000approximating, ammons1997exploiting} to represent the profile. Figure~\ref{fig:cct} shows the structure of a CCT produced by \tool. The internal nodes represent native or Python function calls, and the leaf nodes represents the sampled memory loads or stores.
Logically, each path from a leaf node to the root represents a unique call path. 

As mentioned, Python is a dynamic typing language, and uses meta-data to represent calling context (e.g., the function and file names in string form); therefore, its call stacks are usually substantially larger (in space) than those in static languages. One solution is to build a dictionary to map strings to integer ids but the solution must be signal-free because it needs to interact with the CL-algorithm and PMUs, which is prohibitively complex.

Our crucial observation is that function calls in different threads near the root of a tree usually repeat so unlike solutions appeared in~\cite{chabbi2012deadspy, wen2017redspy, loadspy, chabbi2014call, su2019pinpointing}, which  produce a CCT for each thread/process, \tool{} constructs a single CCT for the entire program execution. In this way, the same function call appearing in different threads is compressed into one node and space complexity is reduced. \tool also implements a lock-free/signal-safe skip-list~\cite{pugh1990skip} to maintain CCT's edges for fast and thread-safe operations. In theory, Skip-list's lookup, insert, and delete operations have $O(\log n)$ time complexity.
In practice,
Skip-list with more layers has higher performance but higher memory overhead. In a CCT, the nodes closer to the root are accessed more frequently. \tool, however,proportionally adjusts the number of layers in the skip-lists at different levels in a CCT to optimize the performance and overhead tradeoffs. It uses more layers to represent the adjacency lists of nodes that are close to the root, and fewer layers to represent those that are close to the leaves.

\subsection{Safeguard}
\tool uses two mechanisms to avoid unexpected errors in Python runtime. It will hibernate if it enters a block of code, interrupting which will cause state corruption in PVM, and will block certain activities from GC if the activities can cause memory issues.

\begin{table*}[!htbp]
    \centering
    \tiny
    \resizebox{\linewidth}{!}{
    \begin{tabular}{||c|c|c||c|c||c|c||}
    \hline
        \multicolumn{3}{||c||}{Program Information} &
        \multicolumn{2}{c||}{Inefficiency} & 
        \multicolumn{2}{c||}{Optimization}  \\
    \hline
        Applications & Library & Problem Code & Category & Pattern  & AS & FS \\
    \hline
    \hline
        \multirow{2}{3em}{Ta~\cite{ta}} & \multirow{2}{2em}{Ta}  & 
        volatily.py(45)/trend.py(536,  &  \multirow{2}{*}{Slice underutilization}  & \multirow{2}{*}{$L$}    & \multirow{2}{2em}{1.1$\times$}  & \multirow{2}{3em}{16.6$\times$} \\
        
        &   & 549, 557, 571, 579)
         &    &    &   &  \\        
    \hline
        NumPyCNN~\cite{numpycnn} & Numpy~\cite{harris2020array, van2011numpy}  & numpycnn.py(161)  & Loop-invariant computation  & $S$  & 1.8$\times$  & 2.04$\times$    \\ 
    \hline
        Census\_main & NumpyWDL~\cite{numpywdl}  & ftrl.py(60)  & Loop-invariant computation &  $S$  & 1.03$\times$  & 1.1$\times$   \\ 
    \hline
        Lasso & Scikit-learn~\cite{scikit-learn}  & least\_angle.py(456, 458)  & Inefficient algorithms & $S$  &  1.2$\times$ & 6.1$\times$ \\ 
    \hline
        \multirow{2}{5em}{IrisData~\cite{irisdata}} & \multirow{2}{3em}{Numpy}  & nn\_backprop.py(222, 228,    &  Slice underutilization \& & \multirow{2}{*}{$L$}  & \multirow{2}{2em}{2$\times$}  & \multirow{2}{2em}{2.02$\times$}   \\
    
        &  & 247, 256, 263, 271, 278)  & API misuse  &    &  &    \\
    \hline
        \multirow{2}{5em}{Network} & Neural-network-  & \multirow{2}{9em}{network.py(103-115)}  & \multirow{2}{6em}{Repeated NFC}   & \multirow{2}{*}{$L$} &  \multirow{2}{2em}{1.03$\times$}  &  \multirow{2}{2em}{1.05$\times$} \\ 
        
        & from-scratch  &  &  &   &   &   \\ 
    \hline 
        Cnn-from-scratch~\cite{cnnscratch} & Numpy  & conv.py(62)  &  Slice underutilization  & $L$  & 2.5$\times$  & 3.9$\times$    \\ 
    \hline
        \multirow{5}{*}{Metaheuristics~\cite{nguyen2019building, nguyen2018resource}} & \multirow{5}{*}{Numpy} & FunctionUtil.py(374)  & API misuse
        & $L$  & 1.4$\times$  & 1.9$\times$    \\
    \cline{3-7}
        &  & FunctionUtil.py(270)  &  Slice underutilization & $L$ &  6.3$\times$  & 27.3$\times$ \\
    \cline{3-7}
        &  & FunctionUtil.py(309, 375)  &  Loop-invariant computation & $S$  & 1.04$\times$  & 1.4$\times$    \\
    \cline{3-7}
        &  & FunctionUtil.py(437) & Repeated NFC & $L$ &  1.02$\times$  & 1.1$\times$ \\

    \cline{3-7}
        &  & EPO.py(40)  & Loop-invariant computation & $S$ & 1.1$\times$  & 1.1$\times$    \\        
    \hline
        LinearRegression~\cite{linearregression} & LinearRegression   & LinearRegression.py(49, 50)  & Repeated NFC  & $L$ & 1.4$\times$  & 1.5$\times$    \\
    \hline
        Pytorch-examples~\cite{pytorch-example} & PyTorch~\cite{paszke2017automatic}  & adam.py:loop(66)  & Loop-invariant computation  & $L$ &  1.02$\times$ & 1.07$\times$     \\ 
    \hline
        Cholesky~\cite{zhou2020harp} & PyTorch  & cholesky.py(76)  & Slice underutilization  & $L$ & 3.2$\times$  & 3.9$\times$    \\
    \hline
        GGNN.pytorch~\cite{ggnn} & PyTorch  & model.py(122, 125)  & Loop-invariant computation & $S$  &  1.03$\times$  &  1.07$\times$   \\
    \hline
        Network-sliming~\cite{Liu_2017_ICCV} & \multirow{2}{*}{Torchvision~\cite{torchvision}}  & \multirow{2}{*}{functional.py(164)}  & \multirow{2}{*}{Slice underutilization} & \multirow{2}{*}{$L$}  & 1.1$\times$  &  1.7$\times$ \\
    \cline{1-1}
    \cline{6-7}
        Pytorch-sliming~\cite{Liu_2017_ICCV} &   &   &   &  & 1.04$\times$ &  1.7$\times$ \\
    \hline
        Fourier-Transform~\cite{fourier} & \multirow{3}{*}{Matplotlib~\cite{Hunter:2007}}  & \multirow{3}{*}{transforms.py(1973)} & \multirow{3}{6em}{Repeated NFC}   & \multirow{3}{*}{$S$} & 1.02$\times$  & 2.8$\times$    \\
    \cline{1-1}
    \cline{6-7}
        Jax~\cite{jax2018github} &   &  &   &   & 1.04$\times$  & 2.8$\times$  \\    
    \cline{1-1}
    \cline{6-7}
        Autograd~\cite{autograd} &   &   &   &   & 1.05$\times$  & 2.8$\times$  \\  
    \hline


    
    \end{tabular}
    }
    \caption{Overview of performance improvement guided by \tool. {\it AS} denotes application-level speedup, {\it FS} denotes function-level speedup, $L$ refers to redundant loads and $S$ refers to redundant stores.}
    \label{table}
\end{table*}

\myparabb{\textbf{Hibernation at function-level.}}
Upon seeing an event (e.g., an instruction is sampled or a redundant memory access is detected), the
PMUs or debug registers use interrupt signals to interact with \tool, which will pause Python's runtime. Error could happen if Python run time is performing certain specific tasks when an interrupt exception is produced. For example, if it is executing memory management APIs, memory error (e.g., segmentation fault) could happen, and if Python is loading native library, deadlock could happen.


\tool maintains a list of functions, inside which \tool needs to be temporarily turned off (i.e., in hibernation mode). To do so, \tool maintains a block list of function, and implements wrappers for each function in the list. Calls to these functions are redirected to the wrapper. The wrapper turns off \tool, executes the original function, and turns on \tool again.  

\noindent{\emph{Dropping events vs. hibernation.}} We sometimes drop an event when it is unwanted (Section~\ref{measurement}). Complex logic can be wired to drop an event at the cost of increased overhead. Here, hibernating \tool is preferred to reduce overhead because no event needs to be kept for a whole block of code.


\myparabb{\textbf{Blocking garbage collector.}}
When Python GC attempts to deallocate the memory that debug registers are tracking, errors could occur. Here, we uses a simple trick to defer garbage collection activities: when \tool monitors memory addresses and it  is within a {\tt PyObject}, it increases the corresponding {\tt PyObject}'s reference, and decreases the reference once the address is evicted. This ensures that memories being tracked will not be deallocated. Converting addresses to {\tt PyObject}'s is done through progressively heavier mechanisms. First, {\tt PyObject}'s exist only in a certain range of the memory so we can easily filter out addresses that do not correspond to {\tt PyObject} (which will not be deallocated by GC). Second, we can attempt to perform a dynamic casting on the address and will succeed if that corresponds to the start of an {\tt PytObject}. This handles most of the cases. Finally, we can perform a full search in the allocator if we still cannot determine whether the address is within a {\tt PyObject}.

%% file: evaluation.tex
\section{Evaluation}
\label{evaluation}
This section studies the effectiveness of \tool{} (e.g., whether it can indeed identify interaction inefficiencies) and its overheads. 

We evaluate \tool{} on a 14-core Intel Xeon E7-4830 v4 machine clocked at 2GHz running Linux 3.10. The machine is equipped with 256 GB of memory and four debug registers. \tool is compiled with {\tt  GCC 6.2.0 -O3}, and CPython (version 3.6) is built with {\tt --enable-shared} flag. \tool subscribes hardware event {\tt MEM\_UOPS\_RETIRED\_ALL\_STORES} for redundant stores detection and {\tt MEM\_UOPS\_RETIRED\_ALL\_LOADS} for redundant loads detection, respectively.

\subsection{Effectiveness}

This section assesses the effectiveness of \tool{}, and the breadth of the interaction inefficiencies problem among influential Python packages. The lack of a public benchmark creates two inter-related challenges: \emph{(i)} determining the codebases to examine inevitably involves human intervention, and \emph{(ii)} most codebases provide a small number of ``hello world'' examples, which have limited test coverage. 

We aim to include all ``reasonably important'' open-source projects and we use only provided sample code for testing. While using only sample code makes inefficiency detection more difficult, this helps us to treat all libraries as uniformly as possible. For each of Numpy, Scikit-learn, and Pytorch, we find all projects in Github that import the library, and sort them by popularity, which gives us three lists of project candidates. Our stopping rule for each list differs and involves human judgement because we find that the popularity of a project may not always reflect its importance (e.g., specialized libraries could be influential, but generally have smaller user bases, and are less popular in Github's rating system). For example, Metaheuristics is important and included in our experiment but it received only 91 ratings at the time we performed evaluation. At the end, we evaluated more than 70 read-world applications, among which there are more projects that import Numpy than the other two libraries.

Indentifying a total of 19 inefficiencies is quite
surprising because these projects are mostly written by professionals,
and the sample codes usually have quite low codebase coverage,
and are usually ``happy paths'' that are highly optimized. The fact that we identify 18 new performance bugs as reported in Table 2, 
indicates that interaction inefficiencies are quite widespreaded.

Table~\ref{table} reports that the optimizations following \tool's optimization guidance lead to 1.02$\times$ to 6.3$\times$ application-level speedup (AS), and 1.05$\times$ to 27.3$\times$ function-level speedup (FS), respectively. According to Amdahl's law, AS approaches FS as the function increasingly dominates the overall execution time.  
For the five inefficiency categories we define in Section~\ref{classification} and which are common in real applications, \tool's superior redundant loads/stores detection proves its effectiveness.


\begin{figure*}[!t]
  \centering
  \subfloat[Redundant Stores Detection]{
  \label{fig:slowdown:store}
  \includegraphics[width=0.492\textwidth]{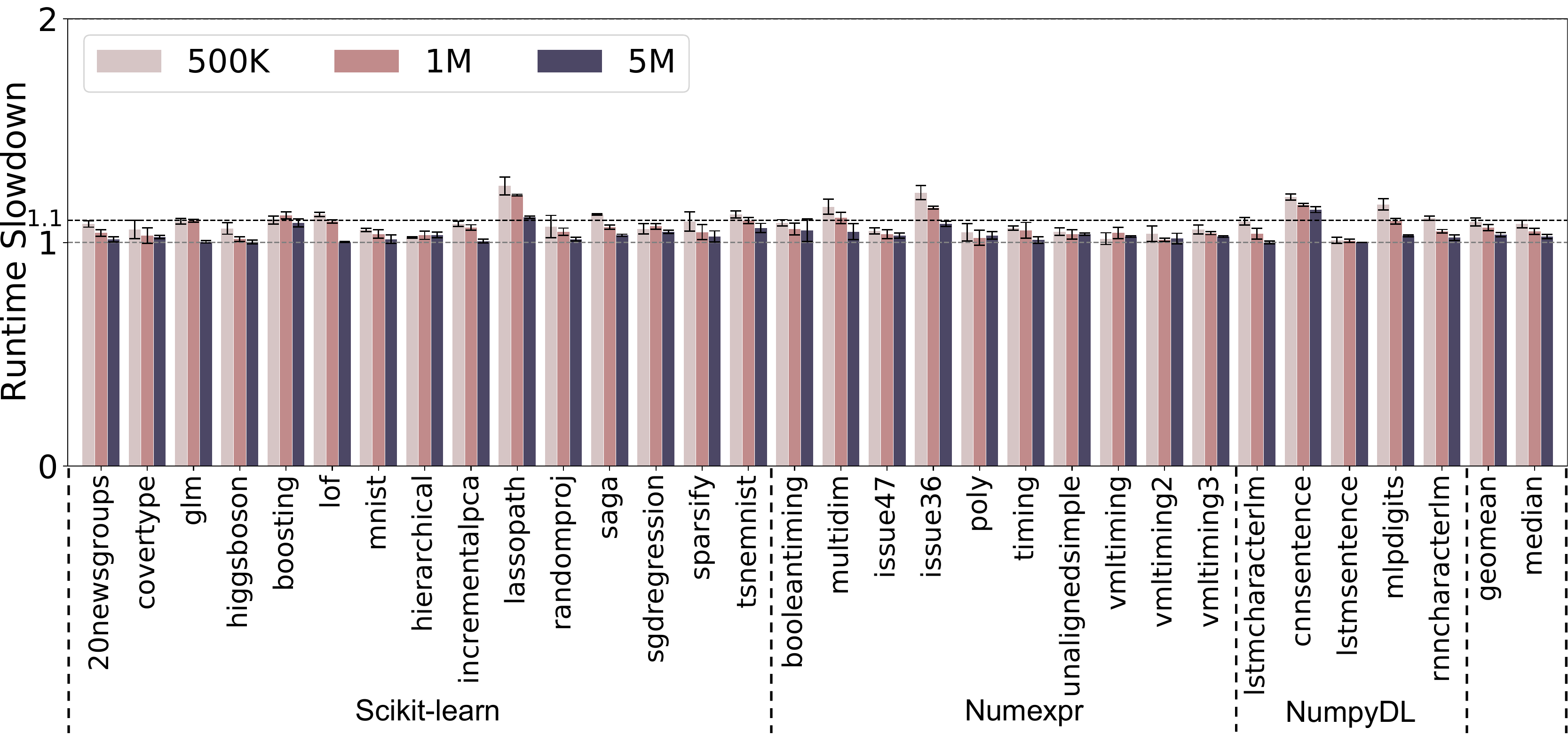}
  }
  \subfloat[Redundant Loads Detection]{
  \label{fig:slowdown:load}
  \includegraphics[width=0.492\textwidth]{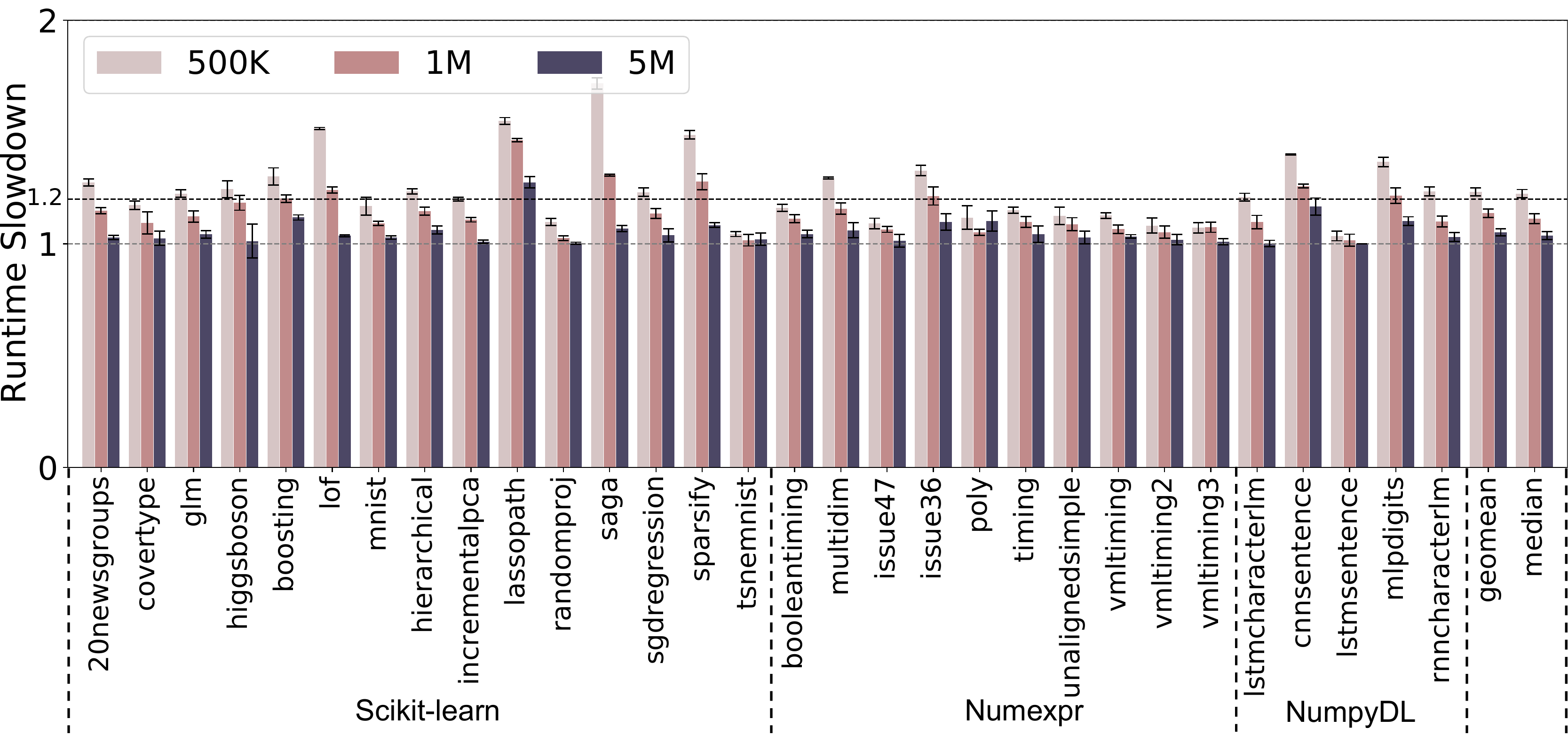}
  }
  \caption{Runtime slowdown of \tool on Scikit-learn, Numexpr, and NumpyDL with sampling rates of 500K, 1M, and 5M. The y-axis denotes slowdown ratio and the x-axis denotes program name.}
  \label{fig:slowdown}
\end{figure*}

\begin{figure*}[!t]
  \centering
  \subfloat[Redundant Stores Detection]{
  \label{fig:memory:store}     
  \includegraphics[width=0.492\textwidth]{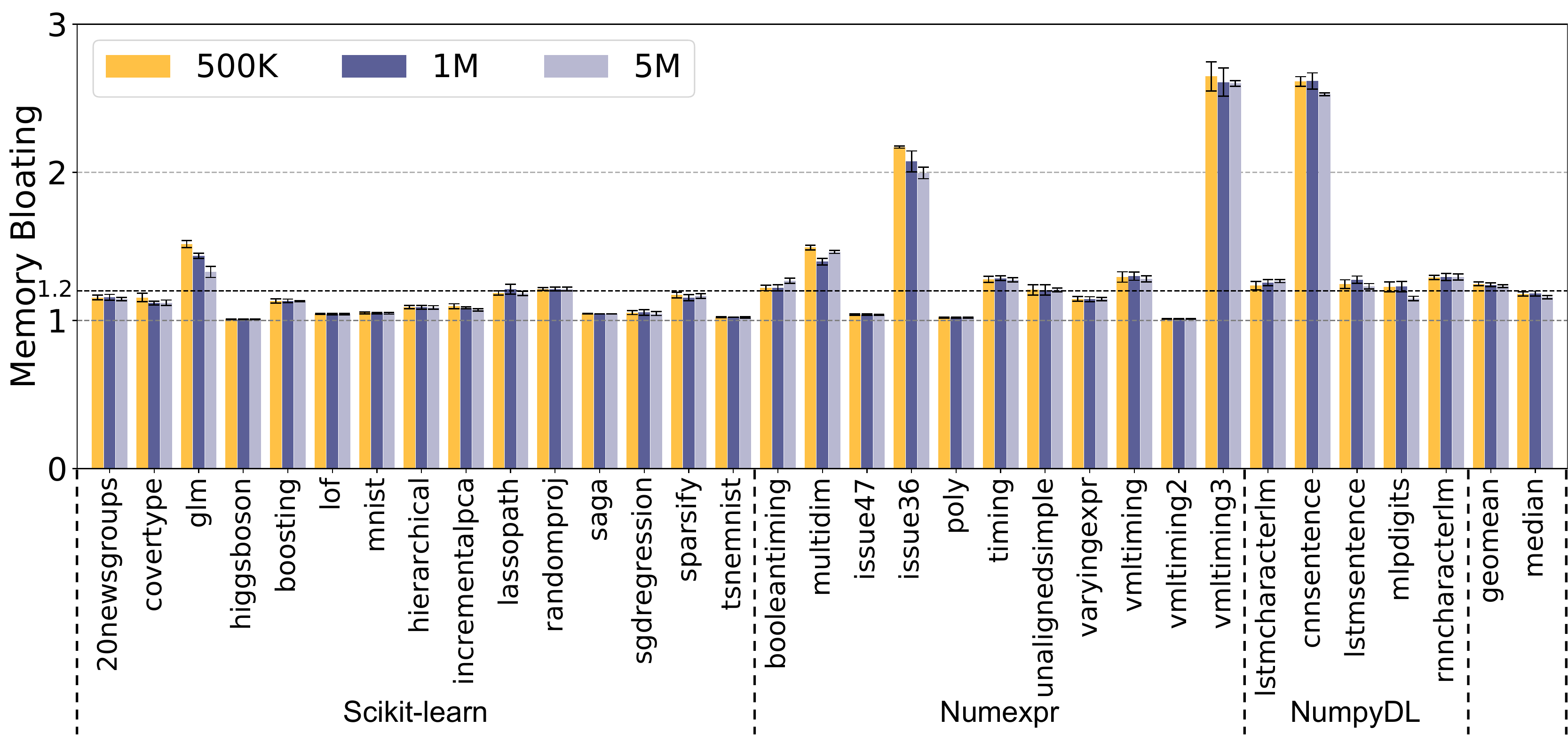}
  }
  \subfloat[Redundant Loads Detection]{ 
  \label{fig:memory:load} 
  \includegraphics[width=0.492\textwidth]{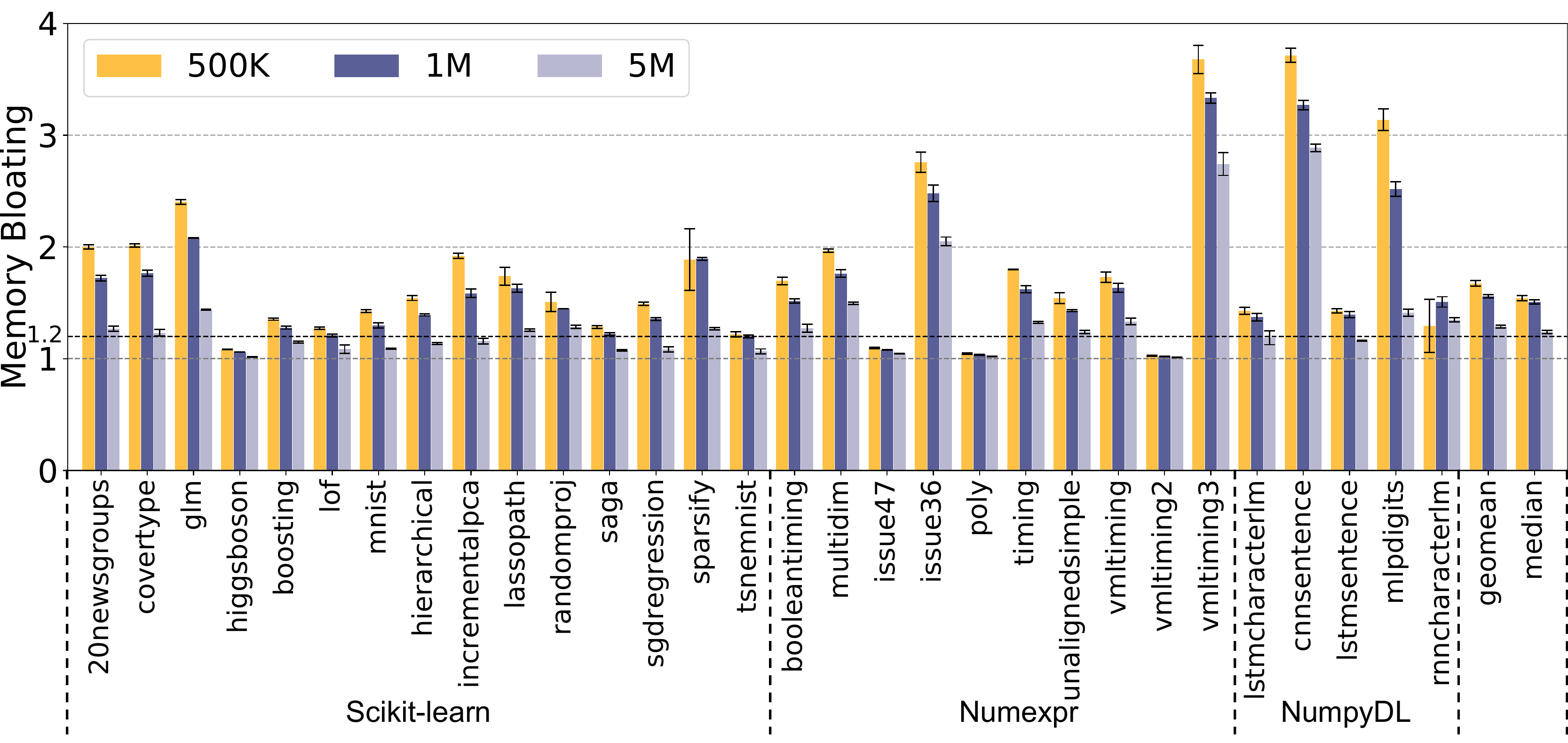}
  }
  \caption{Memory bloating of \tool on Scikit-learn, Numexpr, and NumpyDL with sampling rates of 500K, 1M, and 5M. The y-axis denotes slowdown ratio and the x-axis denotes program name.}
  \label{fig:memory}
\end{figure*}

\subsection{Overhead}

This section reports the runtime slowdown and memory bloating caused by \tool. We measure runtime slowdown by the ratio of program execution time with \tool enabled over its vanilla execution time. Memory bloating shares the same measuring method but with the peak memory usage. 

Since Python does not have standard benchmarks, we evaluate the overhead of \tool on three popular Python applications --- Scikit-learn, Numexpr~\cite{numexpr}, and NumpyDL~\cite{numpydl} which contain benchmark programs from scientific computing, numerical expression and deep learning domains. We report only the first half of the Scikit-learn benchmark due to space limitations, and exclude {\tt varying-expr.py} from Numexpr, {\tt cnn-minist.py} and {\tt mlp-minist.py} from NumpyDL due to large variations in memory consumption, or the runtime errors of vanilla runs {\tt cnn-minist.py} and {\tt mlp-minist.py}.


We run each experiment three times, and report the average overhead. Furthermore, the overhead of \tool is evaluated with three commonly-used sampling rates, 500K, 1M, and 5M.

Figure~\ref{fig:slowdown:store} shows the runtime slowdown of the redundant stores detection. The geo-means are 1.09$\times$, 1.07$\times$, and 1.03$\times$ under the sampling rates of 500K, 1M, and 5M, and the medians are 1.08$\times$, 1.05$\times$, and 1.03$\times$, respectively. Figure~\ref{fig:slowdown:load} shows the runtime slowdown of the redundant loads detection. The geo-means are 1.22$\times$, 1.14$\times$, and 1.05$\times$, under the sampling rates of 500K, 1M, and 5M, and the medians are 1.22$\times$, 1.11$\times$, and 1.04$\times$, respectively. The runtime slowdown drops as sampling rate decreases, because more PMUs samples incur more frequent profiling events, such as inspecting Python runtime, querying the CCT, and arming/disarming watchpoints to/from the debug registers. Redundant loads detection incurs more runtime slowdown compared to redundant stores detection, because programs usually have more loads than stores. Another reason is that \tool sets {\tt RW\_TRAP} for the debug register to monitor memory loads (x86 does not provide trap on read-only facility) which traps on both memory stores and loads. Even though \tool ignores the traps triggered by memory stores, monitoring memory loads still incurs extra overhead.


Figure~\ref{fig:memory:store} shows memory bloating of the redundant stores detection. The geo-means are 1.25$\times$, 1.24$\times$, and 1.23$\times$ under the sampling rates of 500K, 1M, and 5M, and the medians are 1.18$\times$, 1.18$\times$, and 1.16$\times$, respectively. Figure~\ref{fig:memory:load} reports memory bloating of the redundant loads detection. The geo-means are 1.67$\times$, 1.56$\times$, and 1.29$\times$ under the same sampling rates, and the medians are 1.52$\times$, 1.51$\times$, and 1.24$\times$, respectively. Memory bloating shows a similar trend to runtime slowdown with varied sampling rates and between two kinds of inefficiency detection. The extra memory consumption is caused by the larger CCT required for the larger number of unique call paths. {\tt issue36}, {\tt vmltiming2}, and {\tt cnnsentence} suffer the most severe memory bloating due to the small memory required by their vanilla 
runs.  \tool consumes a fixed amount of memory because some static structures are irrelevant to the testing program. Thus, a program has a higher memory bloating ratio if it requires less memory for a vanilla run. {\tt mlpdigits} consumes more memory for redundant loads detection, because {\tt mlpdigits} (a deep learning program) contains a two-level multilayer perceptron (MLP) that has more memory loads than stores. 

Although lower sampling rates reduce overhead, the probability of missing some subtle inefficiencies increases. To achieve a better trade-off between overhead and detecting ability, we empirically select 1M as our sampling rate.

\begin{figure}[t]
\begin{lstlisting}[caption={Interaction inefficiency in CNN-from-Scratch due to slice underutilization.},label=lst:case1_org]
def backprop(self, d_L_d_out, learn_rate):
    d_L_d_filters = np.zeros(self.filters.shape)
    for im_region, i, j in self.iterate_regions(self.last_input):
        for f in range(self.num_filters):
            d_L_d_filters[f] += d_L_d_out[i, j, f] * im_region
\end{lstlisting}
\end{figure}

\begin{figure}[t]
\begin{lstlisting}[caption={Optimized code of Listing~\ref{lst:case1_org}, eliminates inefficiencies by performing slice notation.},label=lst:case1_opt]
def backprop(self, d_L_d_out, learn_rate):
    d_L_d_filters = np.zeros(self.filters.shape)
    for im_region, i, j in self.iterate_regions(self.last_input):
        new_im_region = np.repeat(im_region[np.newaxis,:,:], 8, axis = 0)
        tmp = d_L_d_out[i, j, 0:self.num_filters]
        d_L_d_filters[0:self.num_filters] += tmp[:,None,None] * new_im_region
\end{lstlisting}
\end{figure}

\begin{figure}[t]
	\centering
	\includegraphics[width=\linewidth]{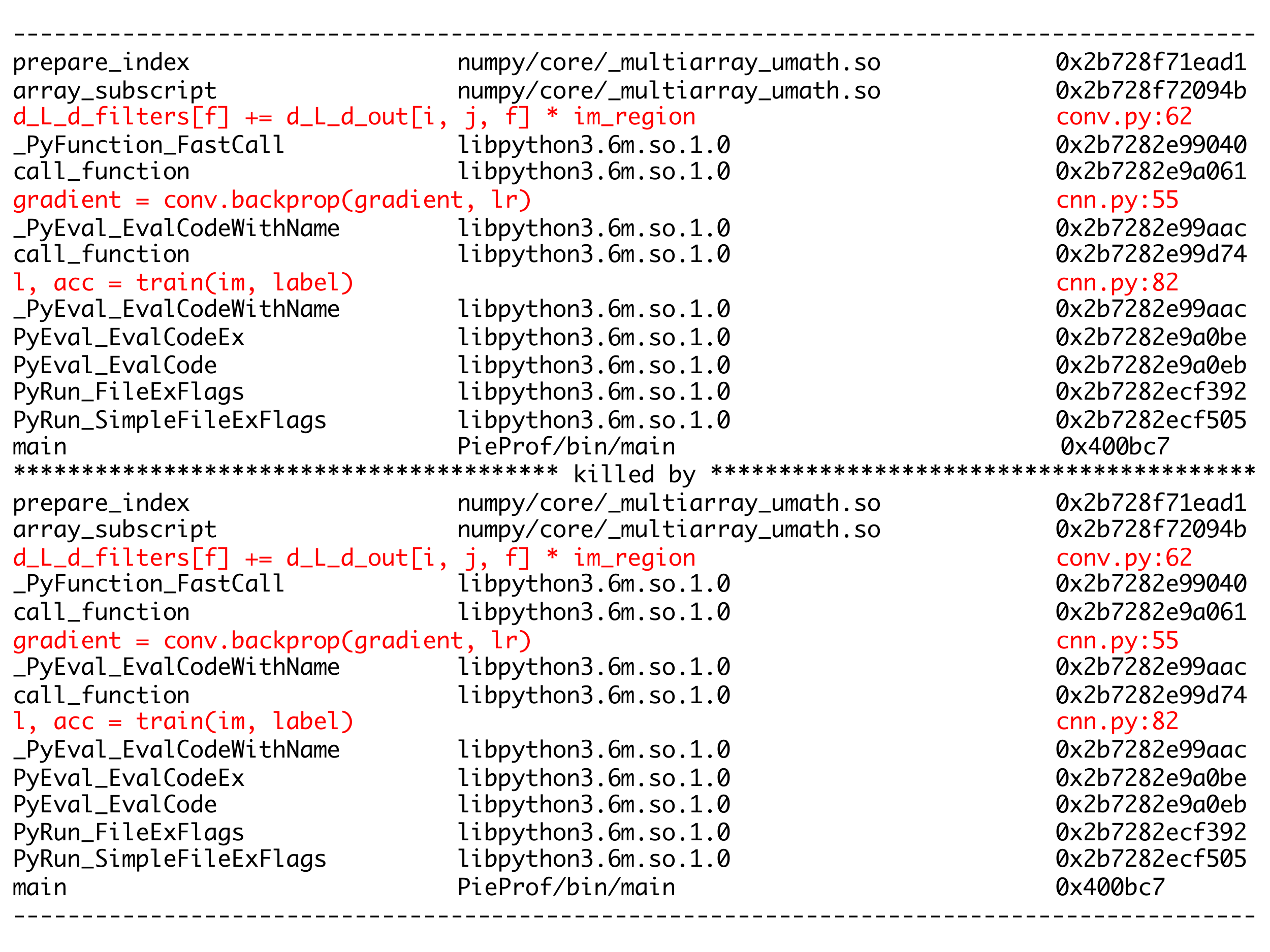}
	\caption{The redundant load pair reported by \tool for Listing~\ref{lst:case1_org}.}
	\label{fig:callpath1}
\end{figure}


\section{Case Studies}
\label{casestudy}

This section discusses our three heuristic case studies. Our primary aim is to demonstrate the superior guidance provided by \tool for inefficiency detection and optimization.

\subsection{CNN-from-Scratch}
CNN-from-Scratch is an educational project that implements a Convolutional Neural Network. The code in Listing~\ref{lst:case1_org} performs tensor computation within a two-level nested loop. {\tt d\_L\_d\_filters} is a 8$\times$3$\times$3 tensor, {\tt d\_L\_d\_out} is a 26$\times$26$\times$8 tensor and {\tt im\_region} is a 3$\times$3 tensor. The inner loop iterates {\tt d\_L\_d\_filters} by its first dimension, iterates {\tt d\_L\_d\_out} by its third dimension. In each iteration of inner loop, {\tt d\_L\_d\_filters[f]} performs as a 3$\times$3 tensor, and {\tt d\_L\_d\_out[i, j, f]} is a number. The computation in line 5 is summarized as a 3$\times$3 vector cumulatively adding the multiplication of a number and a 3$\times$3 vector.

Figure~\ref{fig:callpath1} shows a redundant loads pair reported by \tool. The redundant pair is represented as hybrid call path, and the upper call path is killed by the lower call path. For each native call path, \tool reports the native function name, shared library directory, and the instruction pointer. For each Python call path, it reports the problematic code piece and its location in the source file. In this case, the call path pair reveals that the interaction inefficiency is introduced by line 62 of conv.py (line 5 in Listing~\ref{lst:case1_org}). The call path also shows that the inefficiency caused by \ff \ {\tt prepare\_index({\tt array\_subscript})}, denotes the redundant {\tt []} operations. This inefficiency belongs to the category of slice under-utilization. 

For optimization, we match the dimension of {\tt d\_L\_d\_filters}, {\tt d\_L\_d\_out}, and {\tt im\_region} by expanding the dimension of {\tt im\_region}, and use slice notation to replace the inner loop, as shown in Listing~\ref{lst:case1_opt}. The optimization yields a 3.9$\times$ function-level speedup and 2.5$\times$ application-level speedup.

\begin{figure}[t]
\begin{lstlisting}[caption={Interaction inefficiency in Metaheuristic due to API misuse and loop-invariant computation.},label=lst:case2_org]
def CEC_10(solution=None, problem_size=None, shift=0):
    ...
    for i in range(dim):
        temp = 1
        for j in range(32):
            temp += i * (np.abs(np.power(2, j + 1) * x[i] - round(np.power(2, j + 1) * x[i]))) / np.power(2, j)
        A *= np.power(temp, 10 / np.power(dim, 1.2))
    ...
\end{lstlisting}
\end{figure}

\begin{figure}[t]
\begin{lstlisting}[caption={Optimized code of Listing~\ref{lst:case2_org}, eliminates inefficiencies with an appropriate API and memorization technique.},label=lst:case2_opt]
def CEC_10(solution=None, problem_size=None, shift=0):
    ...
    tmp_dim = 10 / np.power(dim, 1.2)
    for i in range(dim):
        temp = 1
        for j in range(32):
            frac, whole = math.modf(np.power(2, j + 1) * x[i])
            temp += i * np.abs(frac) / np.power(2, j)
        A *= np.power(temp, tmp_dim)
    ...
\end{lstlisting}
\end{figure}


\subsection{Metaheuristics}

\sloppy
Listing~\ref{lst:case2_org} is a code snippet from Metaheuristics. It performs complex numerical computation in a two-level nested loop, where {\tt x} is a Numpy array. \tool reports a redundant loads on line 6, where the code triggers the redundant native function call {\tt array\_multiply} and {\tt LONG\_power}. Guided by this, we observe that {\tt np.abs(np.power(2,j+1)*x[i]} is calculated twice within every iteration, because the code aims to get the computation result's fraction part. To eliminate the redundant computation, we use {\tt math.modf} function to calculate the fraction directly. 

This inefficiency belongs to the category of API misuse in native libraries. \tool also reports redundant stores in line 7 with native function {\tt LONG\_power}. Upon further investigation, we find the result of {\tt np.power(dim, 1.2)} does not change among iterations, which belong to loop-invariant computation. For optimization, we use a local variable to store the result outside the loop and reuse it among iterations. The appropriate usage of API yields 1.4$\times$ application-level speedup and 1.9$\times$ function-level speedup, and eliminating loop invariant computation yields 1.04$\times$ application-level speedup and 1.4$\times$ function-level speedup, respectively.






\subsection{Technical Analysis} Technical Analysis (Ta)~\cite{ta} is a technical analysis Python library. Listing~\ref{lst:ta1} is a problematic code region of Ta, where {\tt adx} and {\tt dx} are two multi-dimension Numpy arrays, and a loop iterates them and performs numerical calculations. 

\tool reports redundant loads in line 6 with native function {\tt array\_subscript}, which denotes the code that suffers from the inefficiency of slice underutilization. Unfortunately, we cannot eliminate the loop because {\tt adx} has computing dependency among the iterations. Therefor, we optimize the access to {\tt dx} with slice notation shown in Listing~\ref{lst:ta2}. Eliminating all similar patterns in Ta yields 1.1 $\times$ application-level speedup and 16.6$\times$ function-level speedup.

\begin{figure}[t]
\begin{lstlisting}[caption={Interaction inefficiency in Ta due to slice underutilization.},label=lst:ta1]
def adx(self) -> pd.Series:
    ...
    adx = np.zeros(len(self._trs))
    tmp = (self._n - 1)/float(self._n)
    for i in range(self._n+1, len(adx)):
        adx[i] = adx[i-1] * tmp + dx[i-1] / float(self._n)
    ...
\end{lstlisting}
\end{figure}

\begin{figure}[t]
\begin{lstlisting}[caption={Optimized code of Listing~\ref{lst:ta1}, eliminates inefficiencies by performing slice notation.},label=lst:ta2]
def adx(self) -> pd.Series:
    ...
    adx = np.zeros(len(self._trs))
    tmp = (self._n - 1)/float(self._n)
    for i in range(self._n+1, len(adx)):
        adx[i] = adx[i-1] * tmp
    adx[self._n+1:len(adx)] += dx[self._n:(len(adx)-1)] / float(self._n)
    ...
\end{lstlisting}
\end{figure}


%% file: conclusion.tex
\section{Threats to Validity}
\label{validity}
The threats mainly exist in applying \tool{} for code optimization. The same optimization for one Python application may show different speedups on different computer architectures. Some optimizations are input-sensitive, and a different profile may demand a different optimization. We use either typical inputs or production inputs of Python applications to ensure that our optimization improves the real execution.  As \tool{} pinpoints inefficiencies and provides optimization guidance, programmers will need to devise a safe optimization for any execution.

\section{Conclusions}
\label{conclusions}

This paper is the {\it first} to study the interaction inefficiencies in complex Python applications. Initial investigation finds that the interaction inefficiencies occur due to the use of native libraries in Python code, which disjoins the high-level code semantics with low-level execution behaviors. By studying a large amount of applications, we are able to assign the interaction inefficiencies to five categories based on their root causes. We extract two common patterns, redundant loads and redundant stores in the execution behaviors across the categories, and design \tool to pinpoint interaction efficiencies by leveraging PMUs and debug registers. \tool cooperates with Python runtime to associate the inefficiencies with Python contexts. With the guidance of \tool, we optimize 17 Python applications, fix 19 interaction inefficiencies, and gain numerous nontrivial speedups.


